\documentclass[a4paper,11pt]{article}
\usepackage{jinstpub} 
\usepackage{rotating} 
\usepackage{subfigure}
\usepackage{siunitx} 
\usepackage{floatrow}
\usepackage{float}
\usepackage{color}
\usepackage[dvipsnames]{xcolor} 
\usepackage{anyfontsize}

\newcommand{\tl}{\tilde{l}}
\newcommand{\tw}{\tilde{w}}
\newcommand{\hh}{\tilde{h}}

\DeclareUnicodeCharacter{025B}{$\varepsilon$}

\usepackage{soul}
\definecolor{lightgreen}{HTML}{B7F774}
\sethlcolor{lightgreen}

\title{\boldmath Permanently magnetized axion-photon conversion surface for direct dark matter searches with BRASS-p}
\usepackage{mathtools}
\author[a]{Le Hoang Nguyen,}
\author[a]{Dieter Horns,}
\author[b]{Andrei Lobanov}

\affiliation[a]{Institut f{\"u}r Experimentalphysik, Universit{\"a}t Hamburg, Hamburg, Germany.}
\affiliation[b]{Max-Planck-Institut f{\"u}r Radioastronomie, Bonn, Germany.}

\emailAdd{le.hoang.nguyen@uni-hamburg.de}

\abstract{A new method was recently proposed for axion dark matter searches which employs scalable, permanently magnetized, flat {\em conversion surface} (or magnetized {\em converter}) generating an electromagnetic signal from dark matter particles passing through them. BRASS-p is an experimental setup which applies this method to direct searches of axion/ALP dark matter in the frequency range from \qtyrange[range-units = single]{12}{18}{\giga\hertz} 
(corresponding to the range of particle mass from \qtyrange[range-units = single]{50}{74}{\micro\electronvolt}.
The conversion surface of BRASS-p will consist of 24 individual panels, each containing a 
$\qtyproduct{483 x 483}{\mm}$ array of high-grade permanent magnets organized in a pattern capable of generating a magnetic field with an average strength of $\approx \qty{0.9}{\tesla}$, for the field component parallel to the surface of the panel. This paper describes the development and optimization of a design for these panels and discusses the resulting sensitivity of BRASS-p measurements to the axion/ALP dark matter.}


\begin{document}
\maketitle
\flushbottom

\section{Introduction}\label{sec:introduction}
    The nature and physical properties of dark matter (DM) remain among the major challenges and central themes of fundamental physics, with a number of possible models and explanations presently being discussed (see \cite{2018RvMP...90d5002B,2024NuPhB100316509S}, for recent reviews). One particular class of these models suggests that DM may be composed of a new scalar particle, the {\em axion} \cite{2009NJPh...11j5008D,2009EPJC...59..557S}. The axion was originally proposed \cite{pecceiCPConservationPresence1977} as an extension to the Standard Model (SM) of particle physics to explain the observed lack of charge-parity (CP) violation in strong interactions \cite{1964PhRvL..13..138C} and the resulting extreme fine tuning of the CP symmetry-breaking parameter $\theta$ (with current experimental limits of $\theta < 10^{-10}$, \cite{2006PhRvL..97m1801B}). The additional, spontaneously broken $U(1)$ symmetry introduced by Peccei and Quinn \cite{pecceiCPConservationPresence1977} (PQ symmetry hereafter) envisaged the parameter $\theta$ to act as a dynamical field, and this field was subsequently associated with a new pseudo-scalar Goldstone boson dubbed the axion \cite{weinbergNewLightBoson1978,Wilczek:1977pj}. At present, the potential relevance of axions for addressing and resolving a number of problems related to DM and cosmology is broadly recognized (e.g., \cite{duffyAxionsDarkMatter2009,2016PhR...643....1M}).
    In these frameworks, the axion is expected to be a low-mass and stable particle and its weak coupling to SM particles is inversely proportional to its symmetry breaking energy scale, $f_\mathrm{a}$. The coupling, $g_{\mathrm{a}\gamma\gamma}$, of the axion to the ordinary photon is given by
    \begin{equation}
        g_{\mathrm{a}\gamma\gamma} = \frac{\alpha}{2\pi f_\mathrm{a} } \left[ \frac{E}{N} -1.92(4)\right]
    \end{equation}
    where $\alpha$ is the fine structure constant and $E/N$ is the ratio between electromagnetic and color anomaly coefficients. The $E/N$ ratio depends on the choice of model for the QCD axion, i.e., the KSVZ, DFSZ1 and DFSZ2  models for which this ratio is 0, 8/3, and 2/3, respectively \cite{DINE1981199,DINE1983137,Zhitnitsky:1980tq,KimPhysRevLett.43.103,SHIFMAN1980493}. In addition to the QCD axion, several extensions of the SM and string theory predict (see \cite{2016PhR...643....1M} and references therein) the existence of a larger set of axion-like particle (ALP) and other weakly interacting sub-eV particles (WISP \cite{2013arXiv1311.0029E}). This includes  dark (or {\em hidden} photons \cite{1982JETP...56..502O,1986PhLB..166..196H}).  All of these additional particles are  considered to be viable dark matter candidates \cite{2012JCAP...06..013A,2014PhLB..732....1J}. 

     In cosmological models that incorporate axions/ALPs as dark matter components, a post-inflationary PQ symmetry breaking scenario is advocated \cite{2008LNP...741...19S,2016PhR...643....1M}. This scenario becomes particularly relevant for axion masses in the range of \qtyrange[range-units = single]{40}{180}{\micro\electronvolt} in order to  match the observed properties of the cold dark matter \cite{Kawasaki:2014sqa,Fleury:2015aca,HiramatsuPRD2012,OHarePhysRevD2022}. 
    
    Numerous experiments for detecting axion dark matter in this particle mass range utilize the {\em haloscope} technique \cite{1983PhRvL..51.1415S,2021RvMP...93a5004S} which relies on axion-to-photon conversion in a resonant microwave cavity placed in a strong magnetic field (e.g., the ADMX 
    \cite{duSearchInvisibleAxion2018}, HAYSTACK \cite{1966IEEEP..54..633R,backesQuantumEnhancedSearch2021}, CAST-CAPP \cite{Adair:2022rtw}, and various cavity-based experiments in South Korea \cite{semertzidisAxionDarkMatter2019,kwonFirstResultsAxion2021}). While providing superb sensitivity to the dark matter signal, all haloscope experiments operate within a limited scanning range of particle mass, $m_a$, near their resonant frequency, $\nu_\mathrm{cav} \approx m_\mathrm{a} c^2$ and thus require a large number of cavity tuning steps to cover that range. Furthermore, the linear
    dimension of a resonant cavity scales approximately with the Compton wavelength of the dark matter particle, and therefore the volume of the cavity becomes progressively smaller for higher particle masses. 
    
    An alternative {\em magnetized dish} approach proposed recently \cite{hornsSearchingWISPyCold2013,2016JCAP...01..005J} overcomes these problems by generating a photonic signal from a conducting {\em converter} surface on which electric currents are excited by the electric field associated with dark matter particles passing through it. This electric field, ${\bf E}_\mathrm{DM}$, is either inherent to the dark matter particles (for dark photons) or induced by the magnetic field above the conducting surface (for axions and ALPs) \cite{hornsSearchingWISPyCold2013}. Interaction of this field with the conducting surface of the converter must satisfy the boundary condition $\textbf{E}_\parallel = 0$, which results in the generation of an outgoing electromagnetic wave $\textbf{E}_\mathrm{a} = \textbf{E}_\mathrm{DM_\parallel}$, propagating perpendicular to the converter surface. For axions and ALP \cite{hornsSearchingWISPyCold2013}:
    \begin{equation}
        \textbf{E}_\mathrm{a}  = \frac{g_{\mathrm{a}\gamma\gamma} \textbf{B}_\parallel \sqrt{2\rho_\mathrm{DM}}}{m_\mathrm{a}} \,, \label{eqn:surfaceE}
    \end{equation}
    where $\textbf{B}_\parallel$ is the parallel component of the external magnetic field near the converter surface, $\rho_\mathrm{DM}$ is the local dark matter density, and $m_a$ is the mass of the axion/ALP particle. The total power $P_\mathrm{a}$ radiated by a converter surface of an area, $A_\mathrm{conv}$, is 
    \begin{equation}
    P_\mathrm{a} = \langle |\textbf{E}_\mathrm{a}|^2\rangle\, A_\mathrm{conv}  = \frac{g_{\mathrm{a}\gamma\gamma}^2 2 \rho_\mathrm{DM}}{m_\mathrm{a}^2} \, \langle B_\parallel\rangle^2\, A_\mathrm{conv} \propto \langle B_\parallel\rangle^2\, A_\mathrm{conv} \,.
    \label{eqn:radiated_power}
    \end{equation}
    An experiment reaching the minimum detectable power, $P_\mathrm{det}$, will therefore be able to probe the axion/ALP dark matter down to the coupling strength of
    \begin{equation}
        g_{a\gamma\gamma} = \left( \frac{1}{\langle B_\parallel\rangle^2 A_\mathrm{conv}}\right)^{1/2} \left( \frac{P_\text{det}}{2 \rho_\mathrm{DM}}\right)^{1/2} m_\text{a}\,,
        \label{eqn:sensitivity}
    \end{equation} 
    and to improve the sensitivity of the experiment, the product $\Phi_\mathrm{A} = \langle B_\parallel\rangle^2 A_\mathrm{conv}$ needs to be maximized, which is an effective equivalent of the volume, $V$, -based factor $\Phi_\mathrm{V} = B^2\,V_\mathrm{cavity}$ applied for performance optimization of microwave cavities used in the haloscope experiments \cite{2013PhRvD..88k5002J}. The traditional approach to achieving this goal employs cooled, superconductive dipole magnets reaching magnetic field strengths of $\sim$\SI{10}{\tesla}, thereby achieving typical factors of $\Phi_\mathrm{A} \sim \SI{40}{\tesla^2\meter^2}$ \cite{2017PhRvL.118i1801C} and $\Phi_\mathrm{V} \sim \SI{5}{\tesla^2\meter^3}$ \cite{2021RScI...92l4502K} in recent and planned experiments. This solution cannot be easily applied to the geometrical setting (spherical or flat converter surface) of the magnetized dish setup, and alternative ways need to be found for producing magnetic field configurations with a sizeable component parallel to the converter surface.

    The Broadband Radiometric Axion/ALPs Search Setup (BRASS) seeks such an alternative in making a flat converter surface itself permanently magnetized and then focusing the outgoing electromagnetic signal with a parabolic reflector onto a heterodyne receiver. This approach significantly increases the area of the converter, so that the full BRASS setup would ultimately use a converter surface of $\approx \SI{100}{\meter^2}$ in area and potentially reach $\Phi_\mathrm{A} \gtrsim \SI{50}{\tesla^2\meter^2}$, depending on the final design of the converter. The prototype demonstrator BRASS-p \cite{2023JCAP...08..077B}, presently in operation at the University of Hamburg, features an unmagnetized converter comprising twenty four \qtyproduct[product-units = power]{50 x 50}{\cm} aluminum panels which provide an effective conversion area of \SI{4.44}{\metre\squared}. The electromagnetic signal produced by the converter is focused with a parabolic reflector of \SI{2.5}{\meter} diameter. The resulting signal is detected with a heterodyne receiver covering the frequency range of \SIrange{12}{18}{\giga\hertz} (particle mass range of  \SIrange{49.63}{74.44}{\micro\electronvolt}). First measurements obtained with BRASS-p have achieved the best sensitivity of all experiments for detecting hidden photon dark matter in this mass range \cite{2023JCAP...08..077B}. The upgraded BRASS-p setup will use the same number of panels, but permanently magnetized, to perform the first broadband searches for axion/ALP dark matter in the same range of particle mass.

This paper describes the current design of the magnetized panels for BRASS-p, discussing their characteristics and performance in generating electromagnetic signals from axion/ALP dark matter. Section~\ref{sec:panel} details the panel design. Section~\ref{sec:em-signal} discusses the outgoing electromagnetic response to dark matter particle passage. Finally, Section~\ref{sec:conclusion} summarizes expectations for BRASS-p measurements using these magnetized panels and explores potential further performance optimization.

\section{Permanently magnetized panel for BRASS-p}
\label{sec:panel}

    A single $50\times 50$\,\SI{}{\cm\squared}, permanently magnetized panel of BRASS-p is composed of  neodymium magnets organized in a two-dimensional pattern such that it maximizes the strength and homogeneity of the magnetic field component parallel to the surface of the panel. Beginning with the pioneering work on using a two-dimensional arrangement of permanent magnets for directional cancellation of magnetic flux \cite{1973ITM.....9..678M}, a number of different arrangements have been introduced, addressing an impressively broad range of problems \cite{2002JMMM..248..441C}. However, none of these designs, including the celebrated Halbach array \cite{1980NucIM.169....1H}, would provide a sufficiently strong and homogeneously distributed magnetic field component oriented parallel to the surface.  This section describes the main aspects and steps in the development of such a design, including the physical properties of individual magnets, the method for calculation of their superposition field, and the process of manufacturing of the first prototype of a permanently magnetized panel for BRASS-p.

    \subsection{Neodymium magnets}
    
    Neodymium (NdFeB) magnets are made of neodymium-iron-boron alloy, forming the $\text{Nd}_2\text{Fe}_{14}\text{B}$ tetragonal crystalline structure. This crystal configuration facilitates magnetization of the material along a specific crystal axis, attributable to four unpaired electrons in the 4f orbital ($4\text{f}^4$) which contribute to the magnetic dipole moment. Consequently, NdFeB magnets exhibit high coercivity, which renders them resistant to high demagnetization field and temperature \cite{10.1063/1.365471,10.1063/1.365469} and thus achieve a remanent magnetic field of up to \SI{1.5}{\tesla}. Neodymium magnets are graded according to their maximum  magnetic flux output per unit volume, and the panel design discussed below employs the NdFeB magnet with the highest commercially available grade (N55, sintered magnet) of NdFeB magnets, reaching a typical remanent field of \SIrange{1.46}{1.52}{\tesla}\footnote{Arnold Magnetic Technologies, https://www.arnoldmagnetics.com/products/neodymium-iron-boron-magnets}.

    \subsection{Magnetostatic field calculation}
    The magnetic field of a single neodymium magnet can be calculated within the magnetostatic approximation. In magnetostatics with no external current, the field intensity ($\bf{H}$) and flux density ($\bf{B}$) are given by 
    \begin{eqnarray}
        \nabla \times {\bf H} = 0 \\
        \nabla \cdot {\bf B}  = 0 \,.
    \end{eqnarray}
    The magnetic flux density, ${\bf B}$, can be expressed through the magnetic vector potential, ${\bf A}_\mathrm{m}$,
    \begin{equation}
        {\bf B} = \nabla\times\bf{A}_\mathrm{m}\,.
    \end{equation}
    For a given magnetization distribution, ${\bf M}({\bf r})$, within a magnetized volume, $V$, with the closed surface $\partial V$, the potential ${\bf A}_\mathrm{m}({\bf r})$ at a position $\bf{r}$ is given by \cite{jackson_classical_1999}
    \begin{equation} 
        \mathbf{A}_\mathrm{m}({\bf r}) = \frac{\mu_0}{4\pi}\int\limits_{V}\frac{\nabla'\times {\bf M}({\bf r}')}{|{\bf r}-{\bf r}'|}\mathrm{d}V'+
                                \frac{\mu_0}{4\pi}\oint\limits_{\partial V}\frac{ {\bf M}({\bf r}') \times {\bf n}' }{|{\bf r}-{\bf r}'|}\mathrm{d}S' \,,
                                \label{eqn:jackson_potential}
    \end{equation}
    where $\mathbf{n'}$ is the outward pointing normal vector at the position of the surface element $\mathrm{d}S'$ and $\mu_0$ denotes the vacuum permeability. For a uniform magnetization distribution, the first term in equation \ref{eqn:jackson_potential} vanishes, and one can integrate equation \ref{eqn:jackson_potential} to obtain a comprehensive analytical solution for a specific shape. 
    
    For the BRASS-p panel, a cuboid shape of individual magnets is selected, which allows for their tiling over the full surface area of the panel, in addition to minimizing the complexity and cost of manufacturing of the individual magnets. The complete analytical solution for a cuboid magnet has been thoroughly examined in prior works by \cite{Yang_1990,Herbert2005} and is reproduced in Appendix \ref{app:analytic_solution} for the system of coordinates used in this paper. In our earlier work \cite{Mangels_2017}, we measured the field of a grade N55 NdFeB magnet cube with the size of $a = \SI{12}{mm}$, using the Hall field probe and confirmed the measurement results with analytic calculations and finite element method (FEM) simulation. 
    
\begin{figure}[!t]
        \centering
        \includegraphics[width=1.0\linewidth]{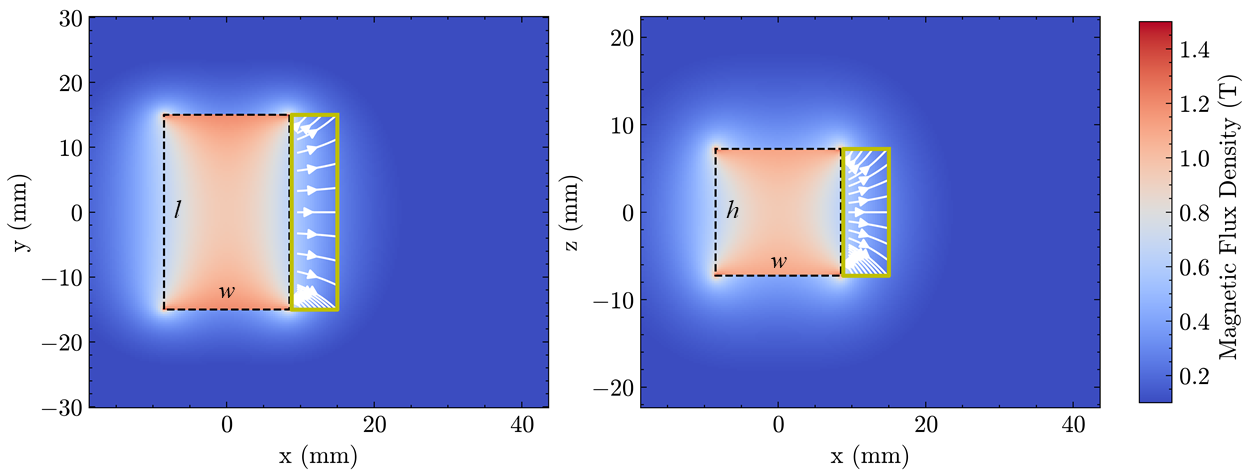}
        \caption{The magnetic field distribution of a single cuboid magnet ($l=\SI{30}{\mm}$, $w = \SI{17}{\mm}$, $h =\SI{14.5}{\mm}$) marked by the dashed box. The geometrical center of the magnet is at the coordinate origin, $(0,0,0)$. Mid-plane distributions $(x,y,0)$ and $(x,0,z)$ of the magnetic field are shown in the left and right panels. The cuboid magnet exhibits a recognizable dipole field pattern, characterized by a gradual decrease in intensity as the distance from the magnet's surface increases. The field strength approaches the nominal maximum value of $\approx \SI{1.4}{\tesla}$ inside the magnet, near its surface. Outside the magnet, the field is both substantially weaker (not exceeding $\approx \SI{0.8}{\tesla}$) and divergent. The light green boxes extended along the x-direction indicate the regions of interest which are used in Sect.~\ref{sec:superpostion_field} for quantifying the performance of the magnetized panel.}
        \label{fig:single_cuboid_field}
    \end{figure}

    In Figure~\ref{fig:single_cuboid_field}, the same calculation approach (using equations \ref{eqn:cuboic1}--\ref{eqn:cuboic3}) is applied for calculating the dipole field of permanent magnets used in the magnetized panel design for BRASS-p (grade N55 NdFeB cuboid magnets, with a remanent field of \SI{1.5}{\tesla} and respective width, length, and height dimensions of $w=\SI{17}{\mm}$, $l = \SI{30}{\mm}$, and $h =\SI{14.5}{\mm}$ respectively, in the Cartesian $x,y,z$ coordinate system). Considering the panel surface to be parallel to the $(x,y)$ plane, the two light green boxes shown in Figure~\ref{fig:single_cuboid_field} outline the volume of space in which the dipole field of the magnet provides the best concentration of the field component, $\langle B_\parallel \rangle$, parallel to that surface. This volume will be at the prime focus of our further analysis of the composite field resulting from multiple magnet arrangements.

    \subsection{Superposition field of multiple permanent magnets} \label{sec:superpostion_field}

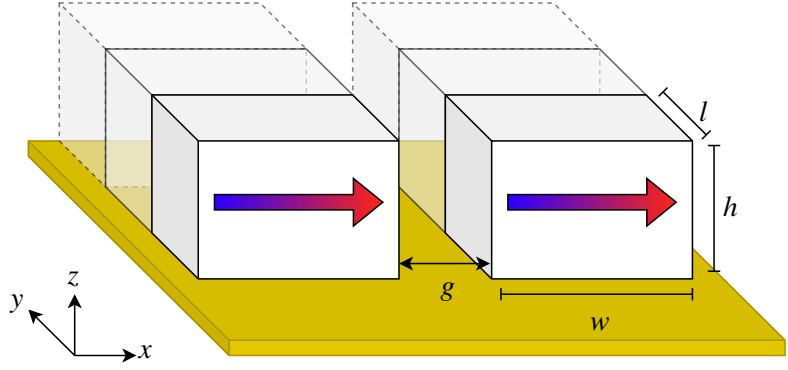
\begin{figure}[!t]
        \centering
        \setlength{\unitlength}{0.05\textwidth}
        \begin{picture}(14,7)
            \put(1,0){\includegraphics[scale =0.8]{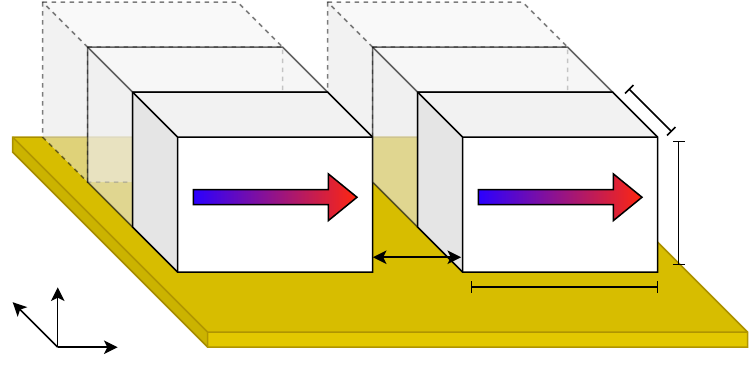}}
            \put(0.9,1.2){$y$}
            \put(3.15,0.3){$x$}
            \put(1.9,1.6){$z$}
            \put(11,0.8){$w$}
            \put(13.3,2.8){$h$}
            \put(12.9,4.45){$l$}
    
            \put(8.4,1.4){$g$}
        \end{picture}
        \caption{
    Conceptual scheme of multiple magnet arrangement for the magnetized panel of BRASS-p. In the air gap $g$ between two cuboid magnets, the superposition field of two opposing magnets produces a strong and homogeneous field component, $\langle B_{\parallel\mathrm{g}}\rangle$, along the $x$ direction, while the influence of more distant magnets in the same row and in the neighboring rows remains limited. With this concept, the inverse Primakoff conversion of axion/ALPs dark matter into detectable photons will occur in the gap as well as close to the surface of the magnets. The latter contribution would be small, as the $\langle B_{\parallel\mathrm{g}}\rangle$ field component is much larger than the $\langle B_{\parallel\mathrm{s}}\rangle \lesssim \SI{0.1}{\tesla}$ generated above the surface of upward faces of individual magnets.
    }
        \label{fig:schematic_panel_design}
    \end{figure}

    Based on the analytical description of magnetic field of a single magnet, we have investigated a number of multiple magnet configurations, attempting to maximize the strength and homogeneity of the $\langle B_\parallel \rangle$ component of the field over the whole panel area. Our attempts to reach these conditions using a continuous surface formed by multiple magnets did not yield satisfactory results, and our present design discussed in this paper uses a different approach. The basic building block of this design, schematically shown in Fig.~\ref{fig:schematic_panel_design}, is a co-linear arrangement of two rows of magnets, with an air gap, $g$, between them and with their magnetization vectors oriented orthogonally to the direction of the gap.
   
\begin{figure}[!t]
        \centering
        \includegraphics[width=1.0\linewidth]{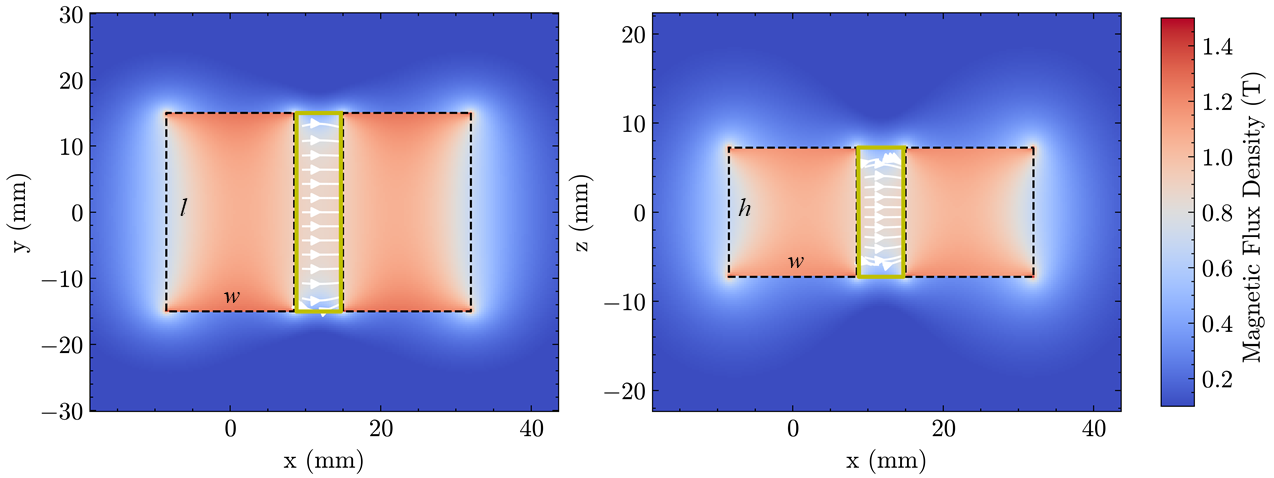}
        \caption{Magnetic field around two cuboid magnets with dimensions of $l = \SI{30}{\mm}, w=\SI{17}{\mm}, h =\SI{14.5}{\mm}$, separated by a gap, $g = \SI{6.5}{\mm}$. The magnetic field in the volume of space between the two magnets is dominated by a component parallel to the $(x,y)$ plane of the magnetic panel. The primary region of interest relevant to BRASS measurements is indicated by light green boxes.}
        \label{fig:double_cuboid_field}
    \end{figure}

\begin{figure}[!t]
        \centering
        \includegraphics[width=0.9\linewidth]{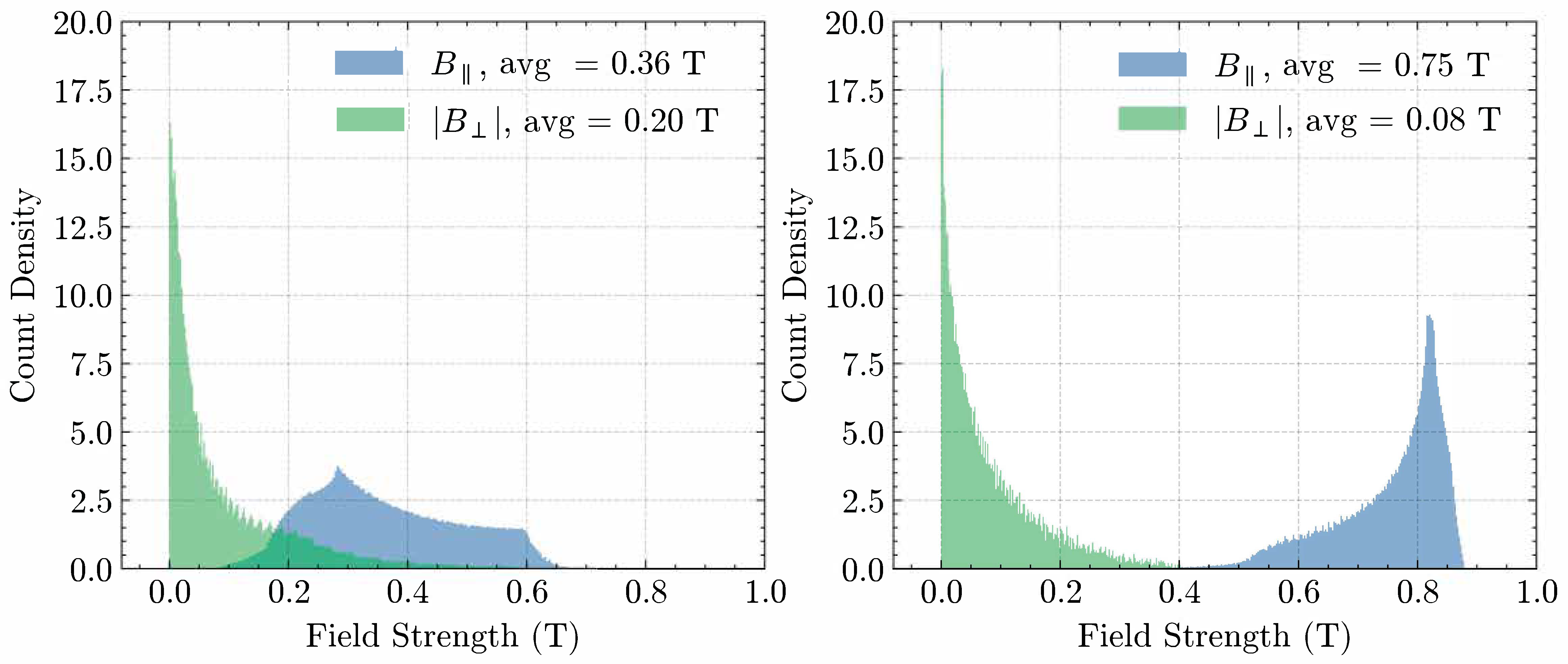}
        \caption{Histograms of strength of the parallel, $B_{\parallel}$, and perpendicular, $|B_{\perp}|$, components of the magnetic field measured in the regions of interests near a single cuboid magnet (left) and between two cuboid magnets (right). With respect to the field properties around a single cuboid magnet, strong suppression of the average $|B_{\perp}|$ and roughly doubling of $B_{\parallel}$ are achieved in the volume of space between two cuboid magnets.}  
        \label{fig:field_strength_histogram}
    \end{figure}

    This arrangement can provide a strongly enhanced and homogeneous $\mathbf{B}_\parallel$ field component within the gap. This is illustrated in Fig.~\ref{fig:double_cuboid_field} showing the field strength (color) and distribution (vector lines) between two cuboid magnets with dimensions of $w=\SI{17}{\mm}$, $l = \SI{30}{\mm}$, $h =\SI{14.5}{\mm}$ and separated by a gap $g = \SI{6.5}{\mm}$. For such a combination of magnets, the dominant field component runs parallel to the $(x,y)$ plane of the magnetic panel, thus satisfying well the experimental needs. This is further demonstrated in Fig.~\ref{fig:field_strength_histogram} which presents histograms of the strength of field components parallel, $B_\parallel$, and perpendicular, $|B_\perp|$, to the surface plane of the magnetic panel, obtained within our regions of interest (the light green boxes in Fig.~\ref{fig:single_cuboid_field} and Fig.~\ref{fig:double_cuboid_field}). It is evident, that $B_{\parallel}$ is strongly enhanced within the volume between two cuboid magnets, while the respective $|B_{\perp}|$ is suppressed. Notably, the average field strength of the parallel component, $\langle B_{\parallel}\rangle= \SI{0.75}{\tesla}$, is roughly two times stronger than what can be achieved in the same volume of space adjacent to a single cuboid magnet. Based on these evaluations, we adopt the magnet arrangement depicted in Fig.~\ref{fig:schematic_panel_design} for the design of magnetic panels for BRASS experiments.

    \subsection{Magnetic panel optimization}\label{sec:magnet_opti}

    According to Eq.~\ref{eqn:sensitivity}, the best performance of a magnetic panel for axion dark matter detection is achieved by maximizing the factor $\Phi_\mathrm{A} = \langle B_\parallel\rangle^2 A_\mathrm{conv}$. Following the discussion in Sect.~\ref{sec:superpostion_field}, we adopt for this purpose the magnet arrangement shown in Fig.~\ref{fig:schematic_panel_design}, with parallel rows of magnets covering the entire panel. The optimization of $\Phi_\mathrm{A}$ for a panel is achieved by varying the geometrical parameters ($l,\, w,\, h$, and $g$)  until the resulting $\Phi_\mathrm{A}$ is maximized. With these considerations, a square panel of $d_\mathrm{p}\times d_\mathrm{p}$ in size would comprise $n_\mathrm{r} = (d_\mathrm{p}+g)/(w+g)$ rows, each containing $n_\mathrm{m} = d_\mathrm{p}/l$ cuboid magnets. With these definitions, $A_\mathrm{eff} = (n_\mathrm{r}-1) d_\mathrm{p}\,g$, provides the effective conversion area of the panel.
    
\begin{figure}[!t]
       \centering
        {\includegraphics[width=0.485\linewidth]{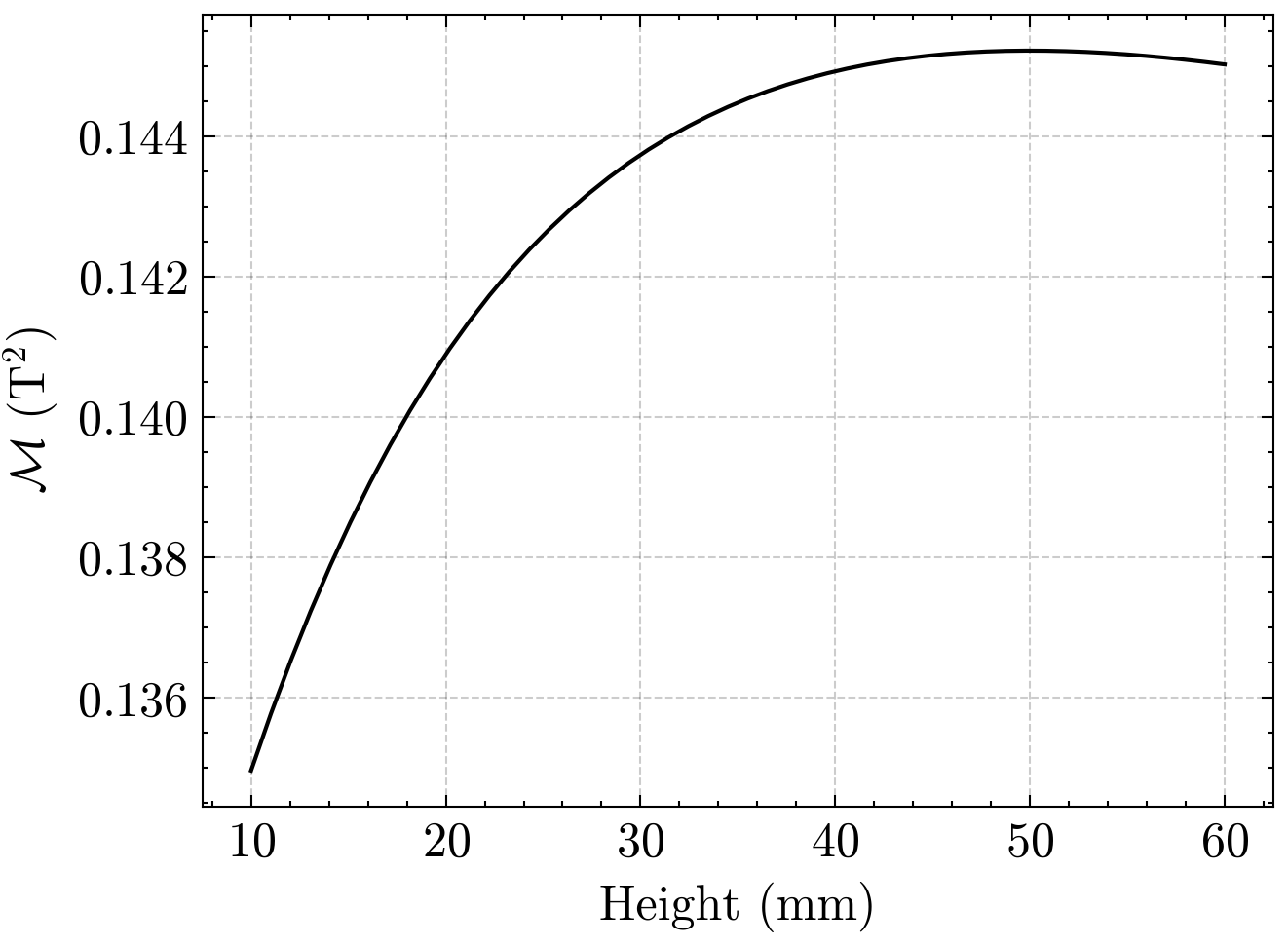}
        \includegraphics[width=0.475\linewidth]{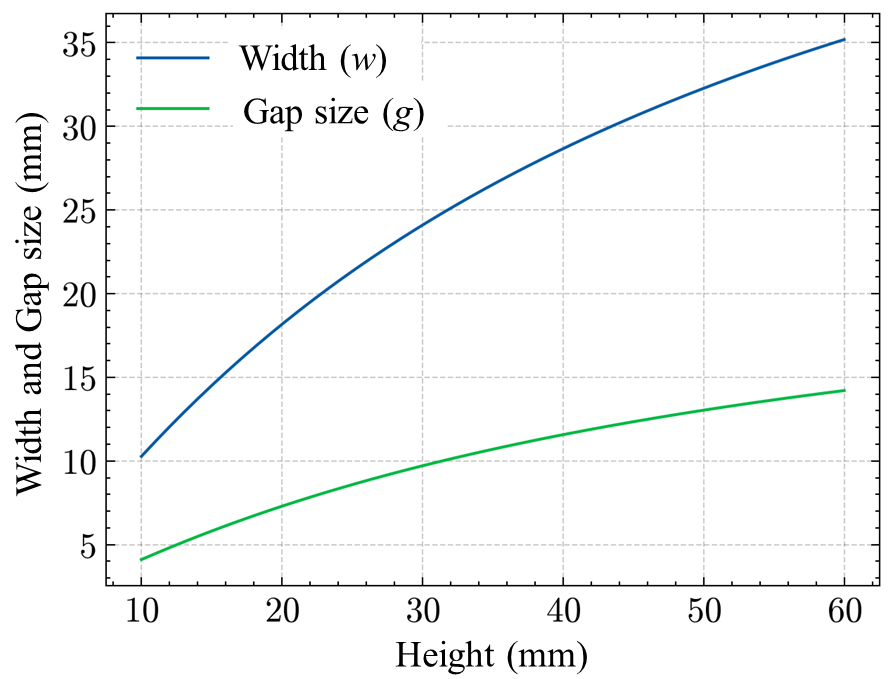}}
        \caption{Optimal values of $\mathcal{M}$ depending on the magnet height, $h$, (left panel) and respective combinations of the magnet width, $w$, and gap size, $g$, between two rows of magnets (right panel). The maximum $\mathcal{M}$ varies only within about 10\%, for $h$ in the range \SIrange{10}{60}{\mm}, thus allowing for a broader selection of magnet and gap parameters delivering comparable performance for BRASS measurements}
        \label{fig:M_optimum_height}
    \end{figure}

    In general, maximizing $\Phi_\mathrm{A}$ should be performed for the full area of the panel, accounting for the $B_\parallel$ field components both within the gap ($|B_{\parallel\mathrm{g}}|$) and above the top surface of the magnets ($|B_{\parallel\mathrm{s}}|$). However, considering that $|B_{\parallel\mathrm{g}}| \gg |B_{\parallel\mathrm{s}}|$ and that field in the gap between two opposing magnets has only a weak (and reciprocated)\footnote{Our calculations show that superposition of the fields from adjacent cuboids results in a uniform increase of the magnetic field strength by approximately 10\,\% over the entire stretch of the gap formed between two strips of cuboid magnets.} effect on the field strength in the adjacent gaps, it suffices to optimize for $\Phi_\mathrm{A}$ over the volume of a single gap, thereby optimizing a reduced (or effective) figure of merit
    \begin{equation}
     \mathcal{M}(w,h,g) = \Phi_\mathrm{A}/A_\mathrm{eff} = \langle B_\parallel (w,h,g)^2 \rangle \,,
    \end{equation}
    with brackets denoting averaging over the gap volume.  The calculation of $\mathcal{M}(w,h,g)$ is made using the method of Broyden, Fletcher, Goldfarb, and Shanno (BFGS) \cite{nocedal2006numerical}. These calculations show that
    $\mathcal{M}(w,h,g)$ depends only weakly the magnet height, $h$, with the maximum values of $\mathcal{M}(w,h,g)$ varying less than 10\,\% over a range of $h$ from \SIrange{10}{60}{\mm} covered by the calculations. This is illustrated in the left panel Fig~\ref{fig:M_optimum_height} showing the change in the maximum $\mathcal{M}$ depending on the magnet height, $h$, in the left panel and the respective combinations of $w$ and $g$ in the right panel. 

\begin{figure}[t!]
    \centering
    \includegraphics[scale=0.6]{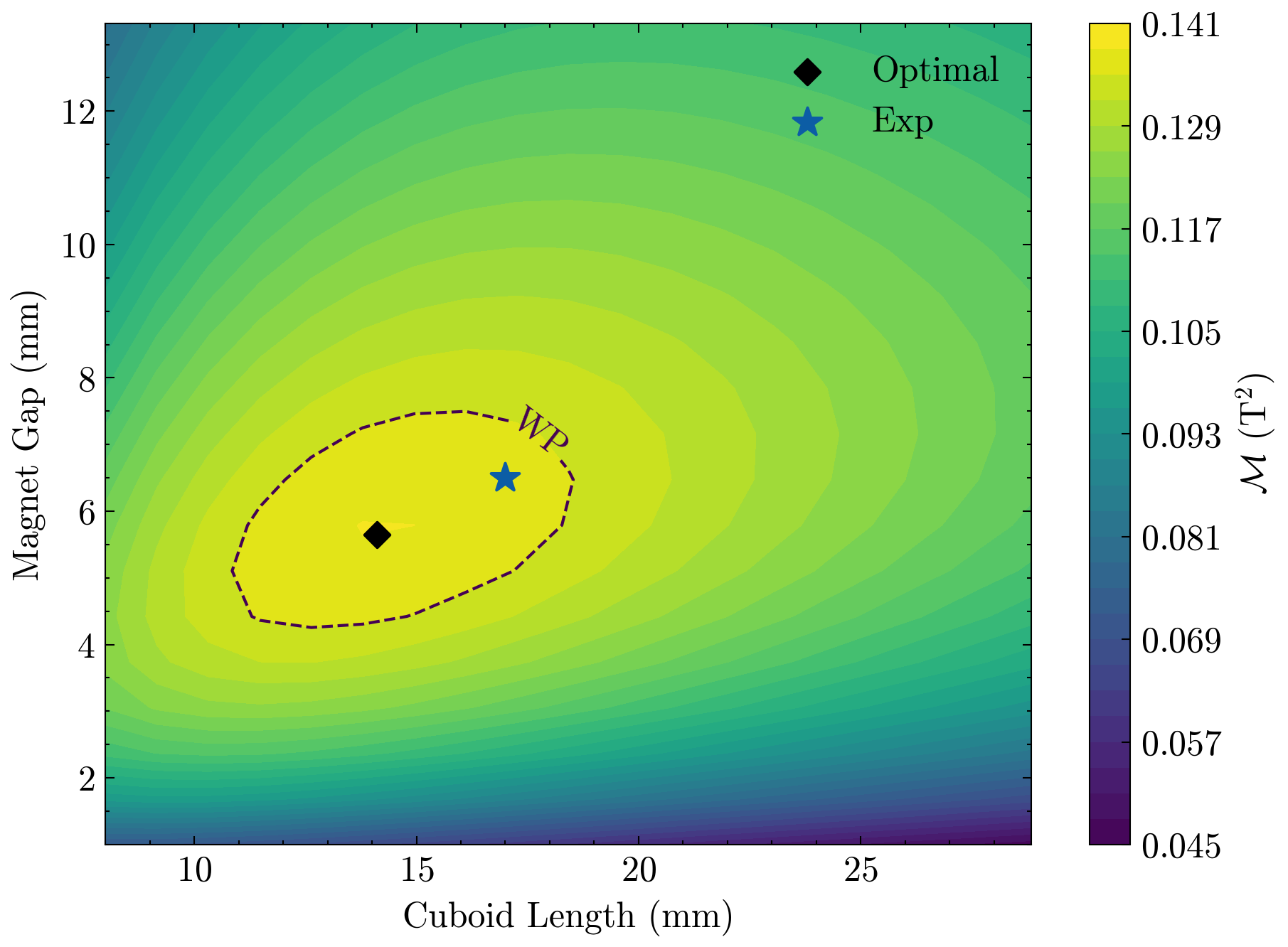}
    \caption{Distribution of the reduced figure of merit $\mathcal{M}$ calculated over a range of $w$ and $g$, for the adopted magnet height of \SI{14.5}{\mm}. The maximum value of $\mathcal{M}=\SI{0.0138}{\tesla^2}$ is obtained for $w = \SI{14.1}{\mm}$ and $g = \SI{5.62}{\mm}$ (diamond marker). The dashed contour marks the 5\% decrease in the $\mathcal{M}$ value which is considered acceptable level of sensitivity loss, thereby allowing for more flexibility in the overall design of the panel. The star indicates the magnet parameters ($w = \SI{17}{\mm}$ and $g = \SI{6.5}{\mm}$) chosen for the panel construction for the BRASS-p prototype experiment.}
    \label{fig:gap_and_length}
    \end{figure}

    Figure~\ref{fig:gap_and_length} shows the distribution of optimal $\mathcal{M}$ obtained for $h= \SI{14.5}{\mm}$ used in the earlier examples. In this distribution, the maximum  $\mathcal{M} = \SI{0.138}{\tesla^2}$ is achieved for $w= \SI{14.1}{\mm}$ and $g= \SI{5.62}{\mm}$, while overall $\mathcal{M}$ remains within about 5\,\% of this maximum within the ranges \SIrange{11}{18}{\mm} in the cuboid length, $w$, and \SIrange{4.5}{7.5}{\mm} in the magnet gap, $g$. Similar ranges of $w$ and $g$ providing nearly optimal values of $\mathcal{M}$ are inferred for other values of $h$ considered. 

    These relations allow for considerable flexibility in designing the magnetic panels, as they indicate that the magnet height and, conversely, the other two geometrical parameters, $w$, and $h$, would not exert strong constraints on the instrumental sensitivity and therefore they can be ultimately determined by other specific constraints such as the panel weight, dimensions and stability, and the longest wavelength, $\lambda_\mathrm{max}$, of measurements, which must satisfy the condition $\lambda_\mathrm{max} \lesssim h/2$ in order to warrant efficient axion-photon conversion at the lowest frequency of measurement. Based on these considerations, the panel design adopted for BRASS measurement will use $h=\SI{14.5}{\mm}$, $w=\SI{17.0}{\mm}$, $g=\SI{6.5}{\mm}$, and $\mathcal{M}\approx\SI{0.14}{\tesla^2}$, which is indicated by the star in Fig.~\ref{fig:gap_and_length}. This design should therefore provide a magnetized converter panel with $A_\mathrm{eff}=\qty{0.057}{\m\squared}$ and $\Phi_\mathrm{A}= \qty[mode=text]{0.028}{T^{2}.m^{2}}$, with the corresponding $\Phi_\mathrm{A}= \qty[mode=text]{0.67}{T^{2}.m^{2}}$ to be realized for the full BRASS-p setup comprising 24 panels.

    \section{Properties of electromagnetic signal generated by a magnetic panel}
    \label{sec:em-signal}
    
    As discussed in Section \ref{sec:introduction}, equation \ref{eqn:surfaceE} describes an outgoing electromagnetic (EM) wave resulting from the conversion of axion/ALPs dark matter  into photons at a magnetized conducting surface. Considering seasonal variations of the laboratory velocity with respect to the dark matter flow, the peak of the EM signal would be located at the velocities between $\approx \SI{170}{\km\,\s^{-1}}$ and $\approx \SI{210}{\km\,\s^{-1}}$ \cite{Nguyen:2019xuh}. This would correspond to the minimum coherence length of $\approx \SI{3.78}{\meter}$ at \SI{18}{\giga\hertz}, and the emitted EM signal would therefore be spatially coherent at all frequencies probed with BRASS-p.

    This condition would give rise to a quasi-planar wavefront far away from the panel, while more specific analysis is required for the near-zone propagation relevant to the BRASS optical setup.
    
    In this regard, there are two constrains that limit the conversion power of this panel 1) Conversion area (design loss): the conversion surface is limited within the gap of the panel, for such a panel shown in figure \ref{fig:initial_design}, the total conversion area is 28\% of the surface area of the panel. 2) Diffraction effect (diffraction loss): if the wavelength of the emitted wave is comparable to the dimensions of the magnet panel, diffraction effects become significant and must be accounted for the total power delivered by the quasi-plane wave. To understand the propagation of converted EM wave from axion/ALPs dark matter, we used FEM simulation and Fourier propagation (Fraunhofer diffraction) method. 

    \subsection{Numerical simulations}
    
    To calculate the properties of electromagnetic signal resulting from dark matter passage through the magnetized conversion surface, we apply the finite element method (FEM) implemented in COMSOL Multiphysics software. To reduce the computation time, the FEM simulation is limited to the $(x,z)$ plane, assuming that the selected plane is separated by $\gtrsim \SI{5}{\milli\meter}$ in the $y$ direction from the panel edge (so that the edge effects on $\mathbf{B}_\parallel$ become negligible; see Fig.~\ref{fig:double_cuboid_field} for a reference). For this setup, the $z$-axis is oriented along the propagation direction of the electromagnetic wave and the $x$-axis is aligned with the orientation of the $\mathbf{B}_\parallel$ component of the magnetic field near the conversion surface.

    The electric field of an outgoing electromagnetic wave generated at the converter surface is given by equation~\ref{eqn:surfaceE}, and based on this equation, two specific models were calculated in order to
    compare the performance of the actual cuboid array to that of an ideal panel with $|\mathbf{B}_\parallel| =\SI{0.65}{\tesla}$ component of the magnetic field, set to match the mean magnetic field of the cuboid array determined in Sect.~\ref{sec:design_and_mechanical_stability}. The strength and spatial distribution of magnetic field of the cuboid magnet array panel are calculated using a FEM simulation for the $l = \SI{30}{\mm}$, $w = \SI{17}{\mm}$, $g = \SI{6.5}{\mm}$ configuration introduced in Sect.~\ref{sec:design_and_mechanical_stability} and illustrated in Fig.~\ref{fig:double_cuboid_field}.

\begin{figure}[t!]
        \centering
        \includegraphics[width=1.0\linewidth]{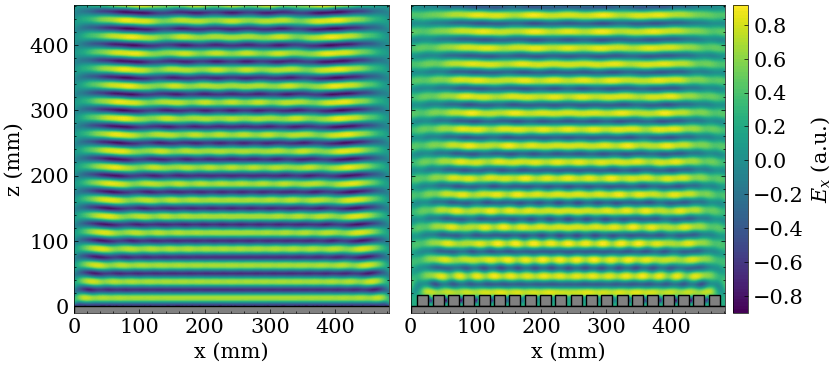}
        \caption{Outward propagation of the electric field ($\mathbf{E}_x$) at a frequency of \SI{12}{\giga\hertz}, generated by the passage of axion/ALP dark matter particles through a conversion surface with a homogeneous (left) and cuboid array (right) magnetic field distribution. In both cases, the electric field distribution approaches the plane wave morphology already at $z\approx\SI{400}{\milli\meter}$.}
        \label{fig:field_propagation_12ghz}
    \end{figure}

\begin{figure}[t!]
        \centering
        \includegraphics[width=1.0\linewidth]{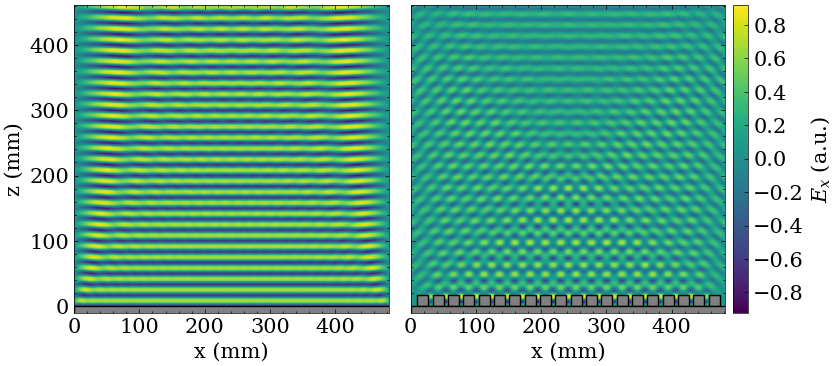}
        \caption{Outward propagation of the electric field ($\mathbf{E}_x$) at a frequency of \SI{18}{\giga\hertz}, generated by the passage of axion/ALP dark matter particles through a conversion surface with a homogeneous (left) and cuboid array (right) magnetic field distribution. In both cases, the electric field distribution approaches the plane wave morphology already at $z\approx\SI{400}{\milli\meter}$.}
        \label{fig:field_propagation_18ghz}
    \end{figure}

The resulting propagation patterns arising at the lowest (\SI{12}{\giga\hertz}) and highest (\SI{18}{\giga\hertz}) operational frequencies of BRASS-p are presented in Figs.~\ref{fig:field_propagation_12ghz}--\ref{fig:field_propagation_18ghz}, respectively. These figures show that the outgoing electromagnetic signal reaches a quasi plane wave morphology already at $z\approx \SI{400}{\milli\meter}$ which is more than 10 times smaller than the \SI{4.8}{\meter} distance between the converter surface and the parabolic reflector of BRASS-p \cite{2023JCAP...08..077B}. This allows us to conclude that the electromagnetic signal produced with the grating structure of the cuboid magnet panel would be successfully concentrated by the parabolic reflector for its subsequent heterodyne detection by the BRASS-p receiving apparatus.

\subsection{Efficiency of the magnetized panel}

To estimate the efficiency of the magnetized panel for dark matter searches with BRASS-p, we use the Fourier propagation method \cite{goodman2005introduction, voelz2011computational} that allows us to fully account for the complex shape of the panel surface and to characterize the evolution of two-dimensional $(x,y)$ distribution of electric field of the outgoing electromagnetic wave as a function of distance from the conversion surface. The Fourier propagation calculations are initialized with the axion-induced electric field at $z = \SI{0}{mm}$, as determined by the external magnetic field structure of the experimental magnet panel. A simplified and conservative assumption of $E_x = const \propto \langle B_\parallel \rangle$ is applied at $z = \SI{0}{mm}$ for every gap location (where $B_x \gtrsim \langle B_\parallel \rangle$), while $E_x \equiv 0$ (reciprocal to $B_x \equiv 0$) is assumed elsewhere near the panel surface. 

\begin{figure}[!t]
        \centering
        \includegraphics[width=0.95\linewidth]{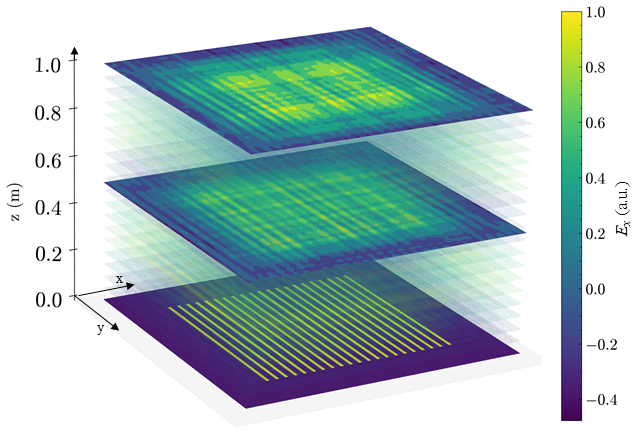}
        \caption{Propagation of the electric field ($\mathbf{E}_x$) at \SI{12}{\giga\hertz} after decomposition in plane waves generated by the passage of axion/ALP dark matter through the magnetized panel. Individual $(x,y)$ planes show the distributions of the field amplitude, $E_x$ at $z=\SI{0}{\meter}$ (depicting the initial condition $E_x = const$ in the gaps between rows of magnets), $z=\SI{0.5}{\meter}$, and $z=\SI{1.0}{\meter}$.}
        \label{fig:fourier_propagation}
    \end{figure}

\begin{figure}[t!]
        \centering
        \includegraphics[width=1\linewidth]{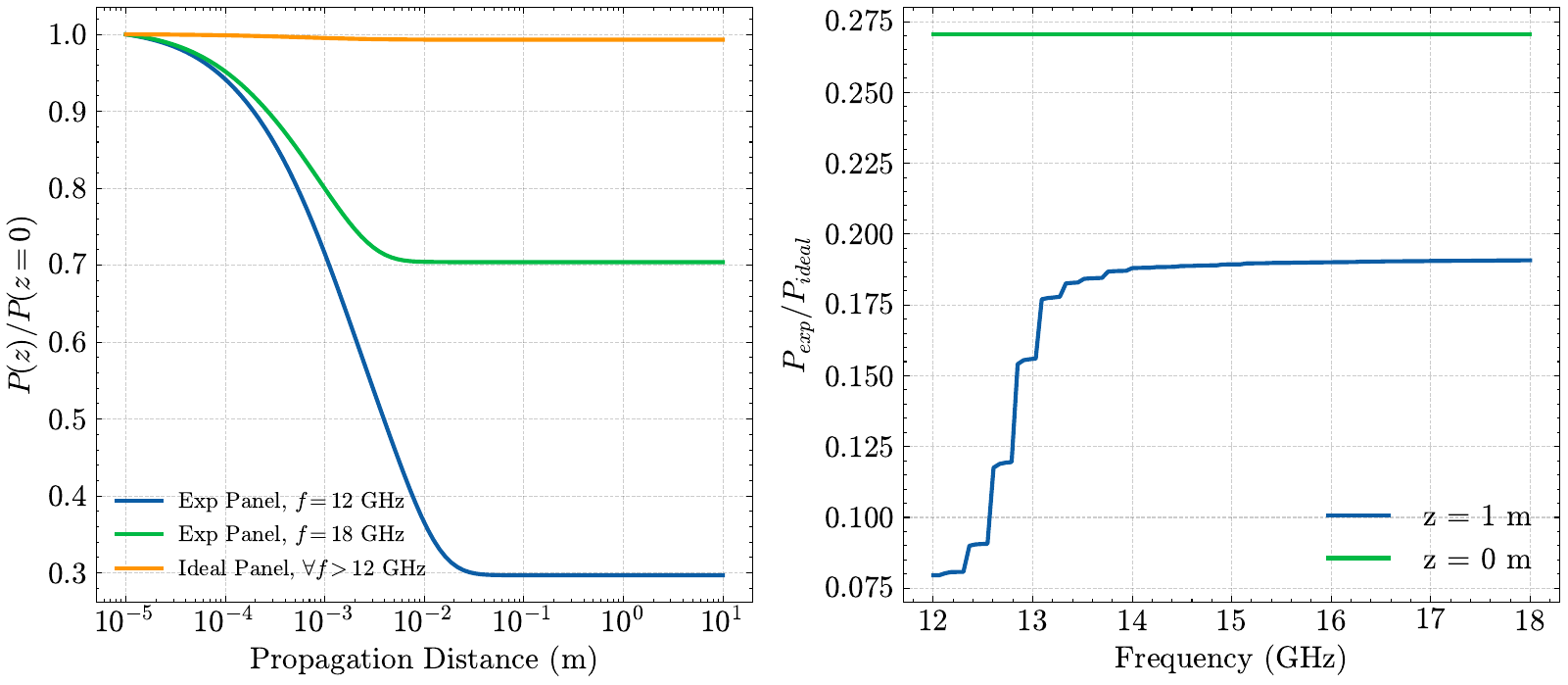}
        \caption{Efficiency of the magnetized converter design for BRASS-p. Left frame shows the diffraction losses of the panel at \SI{12}{\giga\hertz} and \SI{18}{\giga\hertz} as a function of distance from the converter surface, compared to the minimal diffraction loss of the ideal panel. Right frame shows the respective efficiencies of the converter panel at different frequencies and distances from the converter surface. The values at \SI{1.0}{\meter} represent the effective reduction of the converter area due to the partial coverage of the initial field (gaps between the magnet rows) and losses during outward propagation of the electromagnetic signal.}
        \label{fig:fourier_propagation_power}
    \end{figure}

The results of the Fourier propagation calculations are illustrated in Fig.~\ref{fig:fourier_propagation} for the frequency of \SI{12}{\giga\hertz}, showing that the initial sharp-edged distribution of $E_x$ is rapidly evolving into a smoother distribution in the $(x,y)$-plane. Similar calculations are for a range of frequency steps between \SI{12}{\giga\hertz} and \SI{18}{\giga\hertz} and the results are applied for evaluating the efficiency of the magnetized panel at different frequencies. This is illustrated in in Fig.~\ref{fig:fourier_propagation_power}. The left panel of the figure shows the power of the electromagnetic signal, averaged over the $(x,y)$ planes at different $z$ and normalized by the initial power at $z=\SI{0}{\meter}$. Compared to the expectations for a panel with ideal, homogeneous magnetic field, power losses of 30\,\% to 70\,\% are expected in the far field at the operational frequencies of BRASS-p, with higher losses at \SI{12}{\giga\hertz} resulting from stronger diffraction effects at lower frequencies. The resulting estimated efficiencies at $z=\SI{1.0}{\meter}$ are shown in the right panel of Fig.~\ref{fig:fourier_propagation_power} for all frequency steps used in the calculations. These factors represent the effective area of the converter surface and hence must be applied to $A_\mathrm{conv}$ in Eqns.~\ref{eqn:radiated_power}--\ref{eqn:sensitivity}.
%

\subsection{Expected sensitivity of BRASS-p}

\begin{figure}[ht!]
        \setlength{\unitlength}{0.05\textwidth}
        \centering
        \includegraphics[width=0.495\linewidth]{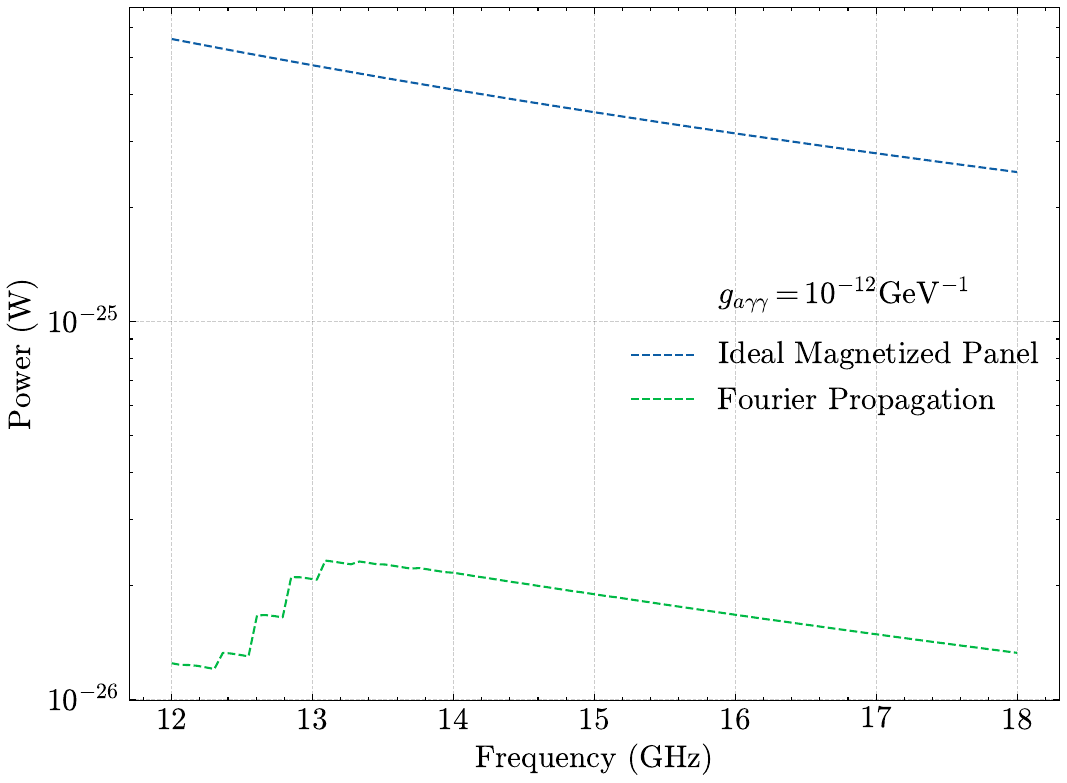}
        \includegraphics[width=0.495\linewidth]{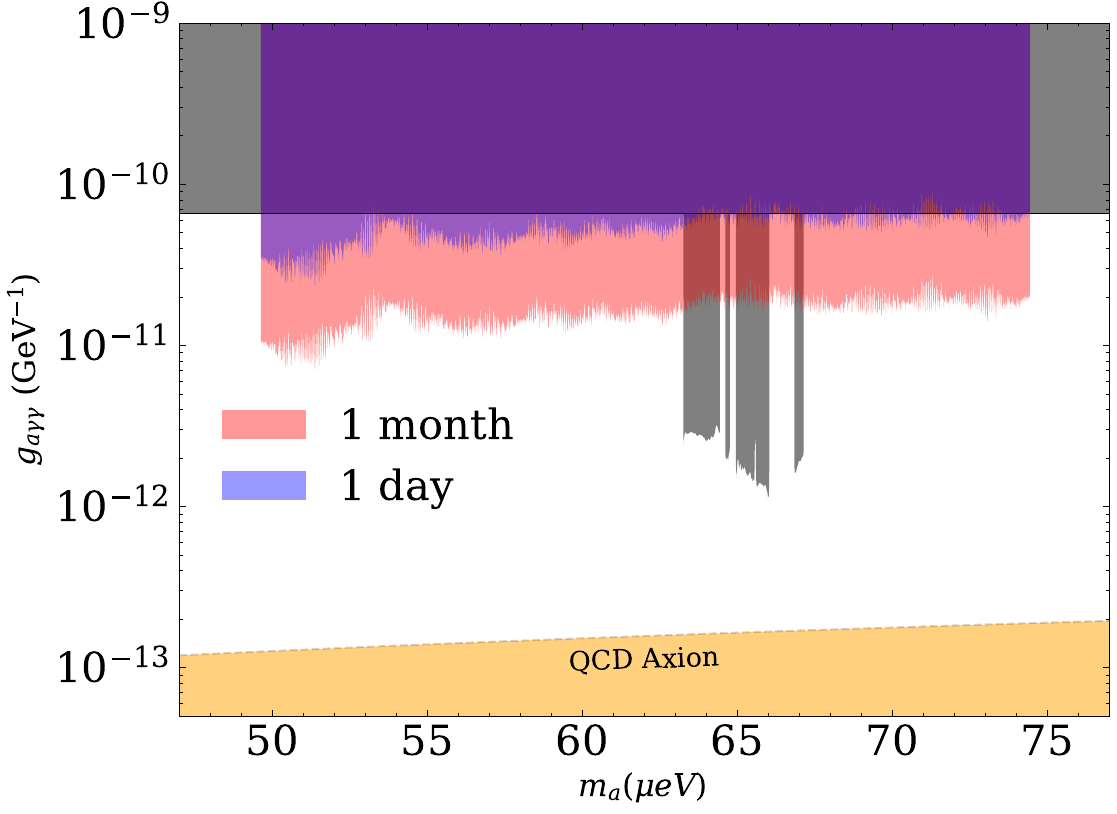}
        \caption{Expected power of the dark matter signal generated by BRASS-p equipped with the magnetized panel described in this work (left panel) and estimated sensitivity of BRASS-p searches for the axion/ALP dark matter (right panel). The grey-shaded area shows ranges of parameter space excluded by earlier experiments. In the \qtyrange[range-units = single]{50}{74}{\micro\electronvolt} range of particle mass, measurements made over a period of one month with the BRASS-p prototype setup providing $\Phi_\mathrm{A}= \qty[mode=text]{0.67}{T^{2}.m^{2}}$ should improve the sensitivity to dark matter by about one order of magnitude and provide useful insights on the ALP dark matter. Reaching the sensitivity required for detecting the dark matter signal from the QCD axions would require increasing the $\Phi_\mathrm{A}$ by a factor of 100 or more.}
        \label{fig:BRASSp_power_sensitivity}
    \end{figure}

The loss factors calculated above can be used for estimating the expected sensitivity of BRASS-p measurements for detecting axion/ALP dark matter, assuming that all 24 aluminum panels used in BRASS-p searches for hidden photon dark matter \cite{2023JCAP...08..077B} are replaced with the magnetized panels. With $\Phi_\mathrm{A}= \qty[mode=text]{0.67}{T^{2}.m^{2}}$ that should be provided by the magnetized converter of BRASS-p, the resulting power expected to be generated by the axion/ALP dark matter is shown in the left panel of Fig.~\ref{fig:BRASSp_power_sensitivity} (assuming the axion-photon coupling $g_{a\gamma\gamma} = 10^{-12}\SI{}{\giga\electronvolt^{-1}}$ and the dark matter density $\rho_\mathrm{DM} = \SI{0.45}{\giga\electronvolt/\cm^{3}}$). Combining this with the current receiver and backend parameters \cite{2023JCAP...08..077B}, the right panel of Fig.~\ref{fig:double_cuboid_field} presents the expected sensitivity to axion/ALP dark matter that can be achieved with BRASS-p measurements with the durations of one day and one month, respectively. The sensitivity estimates demonstrate that the current prototype setup of BRASS-p would be able to improve the existing limits on the axion/ALP dark matter.

\section{Prototype magnetic panel for BRASS}\label{sec:design_and_mechanical_stability}

    Deriving from the panel design developed in this paper, the first prototype of a magnetized panel has been manufactured, considering specifically that the axion dark matter searches planned for BRASS-p should include wavelengths up to $\lambda_\mathrm{max}=\SI{2.5}{\cm}$ ($\nu_\mathrm{min} = \SI{12}{\giga\hertz}$). For an efficient conversion up to this wavelength, the magnet height cannot be smaller than $\approx \SI{12.5}{\mm}$. The other dimensions of the magnet and gap distance are chiefly determined by the constraints on the panel weight and structural stability of the magnet arrangement.

\begin{figure}[t!]
        \centering
        \includegraphics[width=0.7\linewidth]{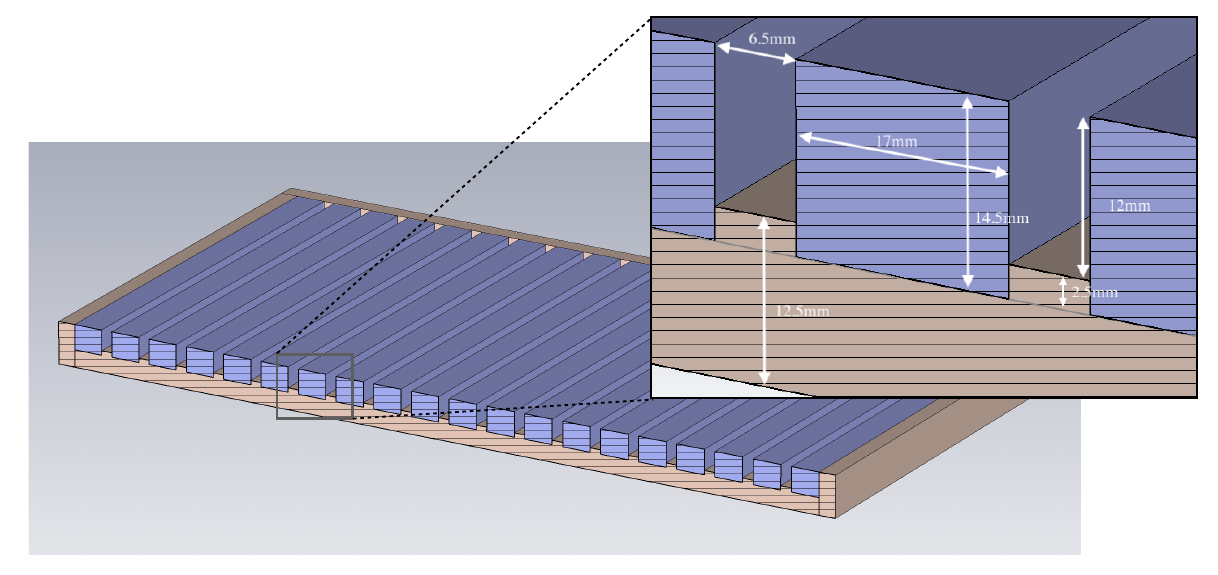}
        \caption{Schematic view of the prototype magnetic panel design that can be fit into the $\SI{500}{} \times \SI{500}{\mm}$ dimensions of BRASS-p panels.}
        \label{fig:initial_design}
    \end{figure}

    The resulting dimensions designed for the prototype panel are illustrated in Fig.~\ref{fig:initial_design}. In order to fit into the present design of BRASS-p \cite{2023JCAP...08..077B}, the panel contains twenty magnetic rows. Each row is composed of cuboid magnets of \SI{17.0}{\mm} in width and \SI{14.5}{\mm} in height (assuming that the individual cuboids are tightly packed along their length dimension, and therefore not requiring to constrain their length). The rows are placed in \SI{2.5}{\mm}--deep grooves cut into an  aluminium support board of \SI{12.5}{\mm} in thickness. The magnetic rows are framed with a \SI{10.0}{\mm}--thick aluminium frame absorbing lateral stresses induced by the magnetic forces between the adjacent rows.  
   
    Attraction force of up to \SI{500}{\newton} also arises between the magnetic rows, and it poses additional difficulty for fixing individual cuboids to their positions within a groove. The magnitude of the attraction force renders it unfeasible to use gluing process for attaching individual cuboid magnets to the board. To resolve this issue, each cuboid magnet is attached to the board with a nylon screw inserted into a bore drilled through the magnet along its $h$-axis and traversing its geometrical center. This procedure minimizes losses and deformations of the magnetic field exerted by the magnet. In view of this modification of the design, the total magnet length $l = \SI{60}{\mm}$ is adopted, to reduce the risk of compromising its structural stability. With these constraints, each magnetic row would contain eight cuboid magnets, providing a magnetic panel of  $\SI{500}{}\times \SI{480}{\square\mm}$ in size. 
    
    \begin{figure}[t!]
        \centering
        \includegraphics[width=0.75\linewidth]{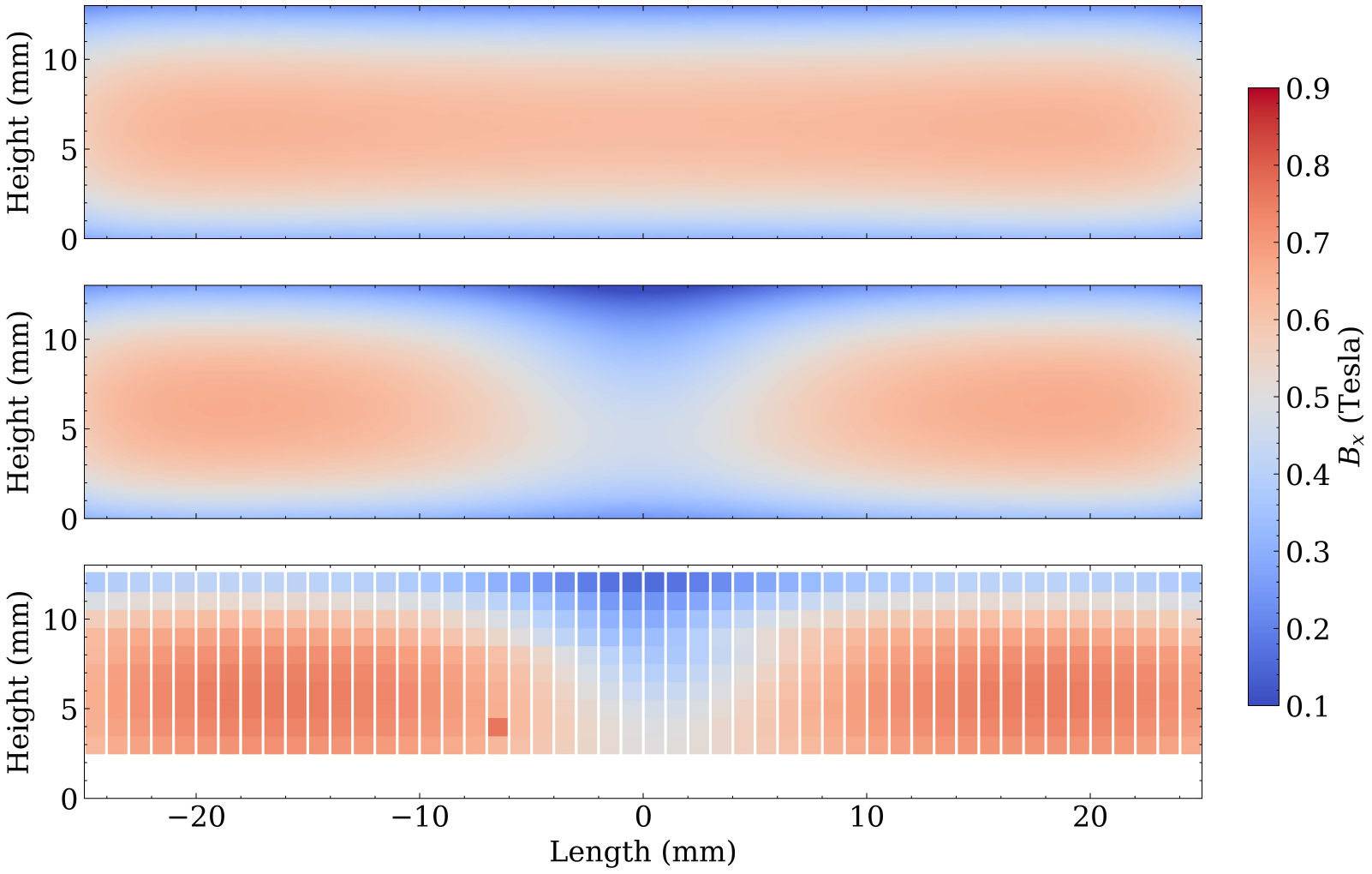}
        \caption{Simulated distributions of the $\mathrm{B_x}$ component of magnetic field of the ideal, intact cuboid (top panel) and experimental, holed cuboid (middle panel), compared to the respective field distribution measured for a cuboid magnet mounted on the prototype magnetic panel.}
        \label{fig:panel_simulation_measurement}
    \end{figure}

    \begin{table}[t!]
        \caption{Simulated and experimental properties of magnetic field exerted by an intact and holed cuboid magnets} \label{tab:cuboid_comparison}
        \begin{tabular}{|l|c|c|c|}\hline
                               &$E_\mathrm{B}$     & $B_\mathrm{x,max}$ & $B_\mathrm{x,mean}$   \\ \hline
        Intact cuboid   & \SI{0.732}{\joule} & \SI{0.830}{\tesla}  &  \SI{0.518}{\tesla}\\ 
        Holed cuboid & \SI{0.681}{\joule}  & \SI{0.848}{\tesla} &   \SI{0.496}{\tesla}\\
        Measured Holed cuboid & N/A  & \SI{0.95(0.15)}{\tesla} &   \SI{0.601 (15)}{\tesla}\\
        
        \hline
        \end{tabular} 
    \end{table}

  The magnetic field reduction resulting from the introduction of a hole in the cuboid magnet has been estimated using finite element modeling (FEM) routine of COMSOL Multiphysics package and verified by using a Hall field probe to measure  \textit{in situ} the magnetic field in the gap of  the panel prototype. Results of the simulations and measurements are compared in Fig.~\ref{fig:panel_simulation_measurement} and Table~\ref{tab:cuboid_comparison} to the magnetic field in the gap between two intact cuboid magnet of the same dimensions. For the holed cuboid, the simulations and measurements both indicate an $\approx 20\,\%$ decrease of the magnetic field strength near the hole and an $\approx 10\,\%$ reduction of the total magnetic energy, $E_\mathrm{B} = (1/2\mu_0) \int_V B^2 \mathrm{d}V$, measured in the volume of interest adjacent to the magnet side facing the opposite magnetic row. The actually measured values of the peak, $B_\mathrm{x,peak} = \SI{0.95 (.02)}{\tesla}$, and mean, $B_\mathrm{x,mean} = \SI{0.65 (.02)}{\tesla}$, magnetic field are larger than the simulated ones, reflecting the contributions from adjacent cuboid magnets in the magnetic row.

\begin{figure}[t!]
        \centering
        \includegraphics[scale=0.7]{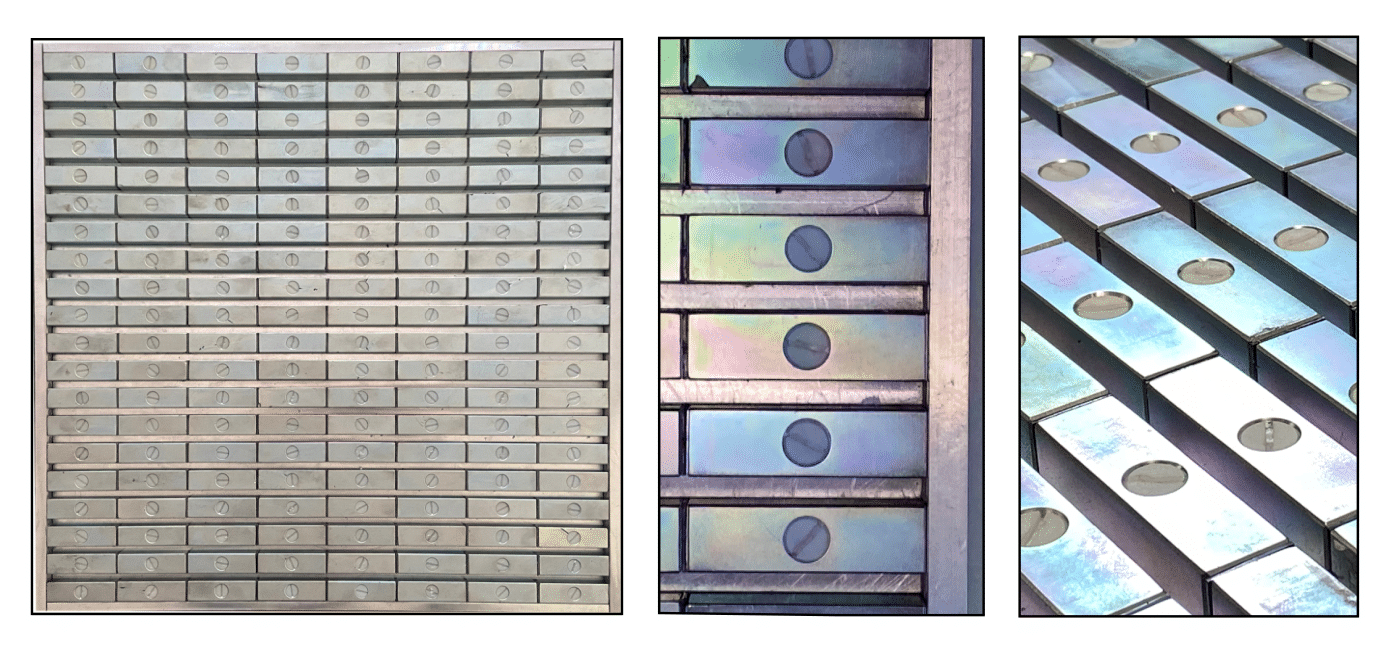}
        \caption{Prototype of permanently magnetized panel for BRASS-p experiment, following the optimized design shown in Fig.~\ref{fig:initial_design}. The panel consists of a total of 160 cuboid magnets (with dimensions $l=\SI{60.0}{\mm}$, $w=\SI{17.0}{\mm}$, $h=\SI{14.5}{\mm}$) arranged in 20 magnetic rows, with each row containing 8 magnets attached along their length dimension, with a thin iron sheets introduced between the magnets in order to prevent damaging their surface during manufacturing of individual magnetic rows). The magnets are attached to the supporting aluminum board with dielectric screws.}  
        \label{fig:real_panel}
    \end{figure}

 This assessment allows us to conclude that the use of nylon screws for attaching individual cuboid magnets to the support board introduces a slight reduction in the overall strength and distribution of the parallel component of magnetic field and hence this design decision can be adopted for the BRASS-p panel. Fig.~\ref{fig:real_panel} shows the actual prototype panel constructed according to the design described above and amended by adding thin iron sheets between adjacent cuboid magnets in each magnetic row in order to prevent their mechanical damage during assembly. The resulting total length of each row is increased to $\SI{481 (1.2)}{\mm}$. The performance parameters of the panel are being presently characterized, and further modifications and improvements of this prototype can be considered in the future.

\section{Conclusions}
\label{sec:conclusion}

The design of a permanently magnetized panel described in this paper provides a viable foundation for performing pioneering BRASS-p searches for the axion/ALP dark matter and, at the same time, represents the initial benchmark to be used for further development and optimization, in preparation for the construction and operation of the full scale BRASS setup. Some potential directions of improvement would include considerations for using new materials for providing even higher magnetic field and development of panel designs optimized for specific frequency bands planned for future BRASS measurements, for instance by modifying the spatial arrangements of the cuboid magnets and using layered dielectric materials for boosting the electromagnetic signal generated at the panel surface.

    \subsection{Stronger cuboid magnets}

     Increasing the magnetic field strength of permanent magnets used for constructing the conversion surface directly enhances the sensitivity of axion/ALP dark matter searches, $ g_{a\gamma\gamma} \propto 1/|\textbf{B}| $. The magnetized panel described above utilizes the highest industrial grade of neodymium, N55, which is readily available and cost-effective. A promising, environmentally friendly alternative to neodymium magnets would be iron nitride ($\alpha''\mathrm{Fe}_{16}\mathrm{N}_2$), which has demonstrated a large remanent field ($B_r > \SI{2.8}{\tesla}$) using abundant raw materials \cite{WANG2020165962}. The use of such magnets would inevitably raise the issue of enhancing the mechanical stability of the panel design, with some of the potential measures in this regard already discussed in Sect.~\ref{sec:design_and_mechanical_stability}. For additional stability of the panel, one can consider filling the gaps between magnet rows with rigid spacers made of polymer materials (e.g., teflon; \cite{Kitai2015}) sufficiently transparent at GHz and THz frequency ranges.

     \subsection{Panel efficiency}
    
   The effective area of the present panel design is chiefly determined by the diffraction losses and, at higher frequencies, would also be affected by the spatial coherence of the outgoing electromagnetic signal. These two aspects would be progressively addressed in future works with regard to specific frequency bands planned for BRASS experiments. Reduction of the diffraction losses could be achieved by increasing the gap between the magnet rows (for instance from the current \SI{6.5}{\mm} to \SI{8}{\mm} with the trade off of 5\% drop in \textit{figure of merit} $\mathcal{M}$; see figure~\ref{fig:gap_and_length}) and, potentially, by simultaneously increasing the height of the cuboid to forestall the reduction in $\mathcal{M}$.

    \subsection{Signal enhancement}
    
     For designs which target specific frequency bands, the conversion power of the magnetized panel can be increased by implementing a layered dielectric coatings, similar to the approach in \cite{majorovitsMADMAXNewRoad2020,DeMiguel:2023nmz}. This layered dielectric coating would be placed inside the gaps of the magnet arrays and can be designed so as to provide a broadband boost factor within a specified frequency range. Zirconia would be the material of choice for such dielectric coating, as it offers one of the highest permittivity among materials available for rapid 3D-print fabrication/prototyping, a low loss tangent at room temperature, and good mechanical properties \cite{DeMiguel:2023nmz}.

\acknowledgments

LHN, DH, and APL acknowledge support by the Deutsche Forschungsgemeinschaft (DFG, German Research Foundation) under Germany’s Excellence Strategy – EXC 2121 „Quantum Universe“ – 390833306.

\appendix
\section{Analytic solution for the magnetic field of a cuboid magnet}\label{app:analytic_solution}
     In the rectangular coordinate system $(x,y,z)$ adopted in this paper, in which the surface of the magnetized panel is parallel to the plane $(x,y)$, the magnetic field of a cuboid magnet with the respective dimensions of $(w,l,h)=(2\tw,2\tl,2\hh)$ and magnetized along the $x$-direction (as shown in Fig.~\ref{fig:schematic_panel_design}), so that $\mathbf{M}=M\,\hat e_x$, 
     can be calculated from (see \cite{Yang_1990}):
\begin{eqnarray}
    B_x &=& - \frac{\mu_0 M}{4\pi}\left( F_1(-x,y,z) + F_1(-x,y,-z) + F_1(-x,-y,z) + F_1(-x,-y,-z)\right.\nonumber \\ 
    &&\left. + F_1(x,y,z) + F_1(x,y,-z)+ F_1(x,-y,z) + F_1(x,-y,-z)\right)  \label{eqn:cuboic1}\\ 
    B_y &=& \frac{\mu_0 M}{4\pi} \ln \frac{F_2(-x,-y,z)F_2(x,y,z)}{F_2(-x,y,z)F_2(x,-y,z)}\label{eqn:cuboic2}\\
    B_z &=& \frac{\mu_0 M}{4\pi} \ln \frac{F_2(-x,-z,y)F_2(x,z,y)}{F_2(-x,z,y)F_2(x,-z,y)}. \label{eqn:cuboic3}
\end{eqnarray}
$F_1$ and $F_2$ are 
\begin{eqnarray}
    \tan F_1(x,y,z) &=& \frac{(\tl+y)(\hh+z)}{(\tw+x)\sqrt{(\tw+x)^2+(\tl+y)^2+(\hh+z)^2}} \\
    F_2(x,y,z) &=& \frac{\sqrt{(\tw+x)^2+(\tl+y)^2+(\hh-z)^2}+\hh-z}{\sqrt{(\tw+x)^2+(\tl+y)^2+(\hh+z)^2}-\hh-z}.
\end{eqnarray}


\bibliographystyle{JHEP}

\bibliography{Reference.bib}
\end{document}